\documentclass{elsart}
\journal{Astroparticle Physics}
\begin{document}

\begin{frontmatter}

\title{Time-integrated supernova neutrino flux from a nearby cluster}

\author{Van T.~Nguyen\thanksref{Columbia}}
\author{and Calvin W. Johnson\corauthref{cor}}
\thanks[Columbia]{Current address: Columbia University, Department of
Physics, 538 W. 120th St New York, NY 10027}

\address{Department of Physics, San Diego State University,
5500 Campanile Drive, San Diego CA 92182-1233, USA}

\corauth[cor]{Corresponding author}

\begin{abstract}
The rate of gravitational collapse (type II) supernovae in our
galaxy is uncertain by a factor of three or more. One way to
determine the galactic supernova rate is through a radiogeochemical
experiment (for example, the molybdenum-technetium experiment) that would
integrate the neutrino flux over several million years. While such
a measurement is designed to integrate the flux over the entire
galaxy, nearby star-forming regions could skew the results. We
model the flux from a recently identified such region, the
Scorpius-Centaurus OB association, and compare with the flux from 
the rest of the galaxy.
\end{abstract}
\end{frontmatter}

\section{Introduction}

Type II supernovae from the gravitational collapse of massive,
short-lived stars are the engines that drive the chemical evolution
of galaxies. While the basic plot of a gravitational collapse
supernova (SN) is known\cite{WHW02}, the details are murky: it is difficult 
to obtain a successful explosion in a computer model, and it is now generally
believed that one must
treat the turbulent hydrodynamics and/or the neutrino transport with
great care in order to produce an outward explosion. The hot,
neutrino-driven neutron wind in the aftermath of a gravitational
collapse SN is a prime candidate for $r$-process nucleosynthesis,
but even that chapter is not finished, inasmuch as it depends
sensitively on entropy, lepton fractions, and neutrino physics\cite{Mey94}.

The famous supernova 1987A \cite{SN1987A} in the Large Magellenic Cloud outside our
galaxy (distance $\sim 50$ kpc) provided us with slightly less than two dozen
neutrino events; with the current generation of neutrino
observatories, should a gravitational collapse SN occur within our
galaxy, we will harvest a wealth of information to help diagnose the
physics of gravitational collapse and subsequent explosion (and
perhaps neutrinos themselves).

Unfortunately, at least for science, galactic SN events are rare.
Various arguments generally suggest the rate of Type II galactic SNe
to be between 1 and 3 per century\cite{VT91}, although some earlier estimates
were as high as 10 per century\cite{BP83}. A  galactic SN has not been visually
observed for nearly four centuries, although we can only visually
observe $\sim 1/6$ of our galaxy; in the 18+ years since SN 1987A,
neutrino observatories have failed to detect any galactic SN, which
should produce hundreds or thousands of events. Knowledge of the
galactic SN rate would help plan for neutrino observatories.

Because the interval between galactic SNe is decades, an alternate
approach to determining the rate is a radiogeochemical experiment\cite{CH82,HJ88}.
A radiogeochemical experiment is a generalization 
of radiochemical experiments such as the chlorine-argon and 
gallium-germanium experiments\cite{Bah89}, but instead of a human-made target,
the parent nuclei are in a naturally 
buried ore body and the daughter is a long-lived but very rare isotope. 
The daughter must be produced 
primarily through neutrino interactions (hence one must be able to 
limit production through fission, cosmic rays, etc.), with  
a half-life short enough 
so no primordial atoms remain but long enough to ``collect'' sufficient data.
As discussed below, and elsewhere, it is possible to identify 
plausible scenarios that will integrate the flux of neutrinos from galactic 
SNe.

Gravitational collapse SNe occur only in massive ($M > 8
M_\mathrm{solar}$) stars with short lives (typically 10-50 My), so
they are associated with star-forming regions. 
If one uses a smooth, exponential parameterization for the 
disk portion of the Milky Way\cite{BS80} one finds a flux 
equivalent to a point source 4.6 kpc distant from Earth\cite{HJ88}. 
But this ignores nearby variations in the SN distribution, which, 
weighted by the inverse of the square of the distance, could have 
a disproportionate effect.

In this paper we consider the effect of a nearby star-forming
region, the Scorpius-Centaurus OB association (Sco-Cen), which is
thought to have had 20 SNs over a period of roughly 10 My
approximately 100-130 pc from Earth\cite{BMC02}.  
In order to understand the effect of the SN neutrino flux from this
star-forming region, we consider the time-dependence of 
the distance to the cluster, through proper motion, 
and the finite width of the clusters.  The focus of this paper is
on the relative contributions from nearby clusters, Sco-Cen in particular,
relative to the neutrino flux from SNe throughout the galaxy.

\section{The Mo-Tc experiment}

The example we consider is the molybdenum-technetium (Mo-Tc)
experiment\cite{CH82,Wo85}; other potential radiogeochemical
experiments are discussed in Ref.~\cite{HJ88}. The Mo-Tc experiment
was originally proposed to measure the long-term solar neutrino
flux, in order to test the hypothesis that the measured low flux of
$^{8}$B solar neutrinos might be caused by nonstandard solar models. 
Such nonstandard solar models have been ruled out through 
helioseismology and detection of neutrino flavor oscillations \cite{BU88}. 
Inverse beta decay induced by neutrino flux on molybdenum ore produces technetium
\begin{equation}
^{98}\mathrm{Mo}+ \nu \rightarrow ^{98}\mathrm{Tc}+e^-.
\label{tc98solar}
\end{equation}
$^{98}$Tc has a half-life of 4.2 My, short enough so that it cannot be
primordial
in origin, but long enough to integrate over the mixing time of the solar core.

Originally the main source of neutrinos considered were 
$^8$B neutrinos from the Sun. It was later realized\cite{HJ88} that
$^{97}$Tc, with a half-life of 2.6 My, could serve as a tracer of
supernova neutrinos through the reaction
\begin{equation}
^{98}\mathrm{Mo}+ \nu \rightarrow ^{97}\mathrm{Tc}+n+e^-.
\label{tc97sn}
\end{equation}
Solar neutrinos are generally too low in energy, however, to knock-out a
neutron and produce $^{97}$Tc through (\ref{tc97sn}), which has a
threshold of 7.28 MeV, but more energetic SN neutrinos can easily do
so. Table~\ref{crosssections} gives the relevant cross-sections as 
computed in \cite{HJ88}.
 (In this
paper we do not consider neutrino flavor oscillations; such
oscillations will certainly change cross-sections\cite{Jo92}, but 
focus only on  relative contributions to supernova 
\textit{fluxes}.)

The larger cross-sections of SN neutrinos allow them to compete with 
solar neutrinos as a source of $^{97}$Tc despite having a lower flux. 
The standard solar model flux of $^8$B solar neutrinos is $\sim 6 \times
10^6$ cm$^{-2}$ s$^{-1}$ \cite{Bah89} (this does not include neutrino 
flavor oscillations, which will reduced this considerably), 
while the flux of galactic SN neutrinos is much
smaller, about
\begin{equation}
R \times N_{58} \times 1.3 \times 10^3 \, \mathrm{cm}^{-2}
\mathrm{s}^{-1}, \label{snflux}
\end{equation}
where $R$ is the rate of galactic gravitational collapse SNe per century, 
somewhere around 1
to 3, and $N_{58}$ is the number of neutrinos emitted per collapse
in units of $10^{58}$, about 0.45 for electron neutrinos. 
(A similar study \cite{PPF04} looks at the neutrino flux from 
thermonuclear burning throughout the galaxy and from 
extragalactic sources.)


This beautiful dichotomy is spoiled somewhat by the
existence of $^{97}$Mo ($9.55\%$ compared to $24.13\%$ for $^{98}$Mo), so that
\begin{equation}
^{97}\mathrm{Mo}+ \nu_{\mathrm solar} \rightarrow
^{97}\mathrm{Tc}+e^-
\end{equation}
can also contribute. The cross-section is not known because there is
no experimental data on the Gamow-Teller strengths, but a reasonable
scaling argument \cite{HJ88}  can be used to derive the $^{97}$Mo cross-sections 
in Table \ref{crosssections}; one would need to subtract off the 
solar contribution in order to measure the galactic SN rate.  (An attempt to
mount the Mo-Tc experiment
was begun by K. Wolfsberg and collaborators in the 1980's, but the
original experiment was never completed due to funding limitations
\cite{Haxton}.  We note, however, that 
the Henderson molybdenum mine in Colorado, the site of the original Mo-Tc experiment,
is still active and is at the time of this writing a candidate for 
siting the National Underground Science Laboratory, and it is not impossible 
that the Mo-Tc experiment could be resurrected.)

\section{Time-dependence and nearby events}

Sco-Cen can be divided into through subgroups: Lower Centarus Crux
(LCC), Upper Centaurus Lupus (UCL), and Upper Scorpius (US), each of whose
distance from the Earth varied between 100 and 240 pc over the past
10 My\cite{Maiz01,BMC02} due to proper motion. Each of the subgroups have a
finite window over which SNe could have occurred: 3 My for US, 7 My
for LCC, and 10 My for UCL.  The SN rate in \textit{each} 
subgroup is assumed to be approximately 1 per My\cite{Maiz01,BMC02}. 

In order to account for the time-dependence of the neutrino flux, we
note we must solve
\begin{equation}
\frac{dN}{dt} = -\lambda N(t) + \sigma \Phi(t),
\label{diffeq}
\end{equation}
where $N(t)$ is the number of daughter nuclei, here $^{97}$Tc,
$\lambda$ is the decay constant, $\sigma$ is the reaction cross-section, and
$\Phi(t)$ is the time-dependent flux. From the data in
Ref.~\cite{BMC02} we approximated $\Phi$ as a quadratic in $t$; this
can be solved analytically. We assumed the rate of SNe is
constant in time.

The most useful way to present the results is to compute an
effective distance $r_\mathrm{eff}$; the effect of the
time-integrated flux for the cluster in motion is the same as a
\textit{static} cluster fixed at the distance $r_\mathrm{eff}$.

We found that the effective distance $r_\mathrm{eff}$ can be approximated by
the
following simple prescription:
\begin{equation}
\frac{1}{4 \pi r^2_\mathrm{eff}} = \langle \Phi \rangle = \frac{
\int_{-T}^0 \Phi(t) e^{-\lambda t} dt }{\int_{-T}^0  e^{-\lambda t} dt}
\end{equation}
where $T$ is the time before the present that SNe began in the subgroup; 
this formulation, which is simpler than solving Eq.~(\ref{diffeq}),  
agrees with such solutions to within a fraction of a percent. 
Table II gives the average fluxes and $r_\mathrm{eff}$ for each subgroup studied. 
The total average flux from Sco-Cen is then the sum of the fluxes in the 
middle column, 450 cm$^{-2}$ s$^{-1}$.

\section{Finite-size corrections and extra-near events}

We also considered the fact that the star-forming subgroups have a
finite width, on the order of 20-30 pc.  By considering a sphere of
uniform density with a center at a distance $r$ from Earth and with rms radius $\sigma$,
we find that the flux from the cluster is
\begin{equation}
\Phi = \frac{1}{4\pi r^2} \left( 1+ \frac{\sigma^2}{3r^2}\right).
\end{equation}
Monte Carlo simulations with a Gaussian distribution but 
the same rms radius agreed with this correction term. Thus 
the correction due to finite width is relatively small. 

We further took the opportunity of our Monte Carlo simulation to 
investigate the question of extra-near SN events. 
Anomalous deposits of $^{60}$Fe in deep-sea crust have been linked 
to a nearby (a few 10 pc) SN within the last few My \cite{Knie99,Knie04}.
Ben\'itez, Ma\'iz-Apell\'aniz, and Canelles \cite{BMC02} suggest this 
anomaly could be explained by a SN in Sco-Cen; 
taking a cluster at 100 pc with a Gaussian with an rms width of 
25-30 pc, they argue for a SN as close as 40 pc from Earth ($2\sigma$ lower
limit). Unfortunately this argument is based upon a one-dimensional 
Gaussian distribution and is wrong for a \textit{three-dimensional}
distribution. In our Monte Carlo simulations, we found relatively few simulation
runs that placed a SN at about 40 pc; in fact only a fraction of a percent 
of any given simulation with 20 SNe gives an event closer than 50 pc. 
Increasing the rms width of the cluster to 40 pc still only yields a 
near-Earth ( $< 50$ pc) event only $12\%$ of the time. Of course,
such simulations depend sensitively on the tail of the distribution,
which might be non-Gaussian. Nonetheless, while Sco-Cen 
is not ruled out as a candidate source for the $^{60}$Fe anomaly, the statistical linkage is 
weaker than previously argued. Conversely, one could possibly 
use $^{60}$Fe and other radioisotopes to constrain nearby events \cite{FHE05} 
and sharpen our modeling of the flux from such events.

\section{Conclusions}

We have considered the potential effect  a nearby star-forming
region, Sco-Cen, would have on radiogeochemical measurements of the
galactic supernova flux, in particular the Mo-Tc experiment.
Although we performed detailed calculations, we found that both the
time-dependence of the flux due to proper motion and the finite
width of a region can be calculated through simple, effective prescriptions 
which could be applied to any cluster and any radiogeochemical experiment
of interest.  The flux from Sco-Cen
is roughly the same as that from throughout the galaxy \textit{if}
the galactic SN rate is 1 per century; this``background'' could
limit the ability of a radiogeochemical experiment to measure the
overall rate.  At higher galactic rates the Sco-Cen contribution
will be relatively small.

\begin{ack}

The U.S.~Department of Energy supported this investigation through
grant DE-FG02-96ER40985.  VTN gratefully acknowledges support from
the SDSU McNair Scholars program and from NIH SDSU MARC grant \#5
T34 GM08303.
\end{ack}

\vfill
\eject

\begin{table}
\caption{Neutrino capture cross-sections for various sources; taken 
from \cite{HJ88}  \label{crosssections}}
\begin{tabular}{lll}
Reaction  & $\nu$ source & $\sigma$ ( $10^{-42}$ cm$^2$)  \\
\hline
$^{98}$Mo($\nu,e^-$)$^{98}$Tc & Solar $^8$B & 3 \\
$^{98}$Mo($\nu,e^-$)$^{98}$Tc & SN  & 50  \\
$^{98}$Mo($\nu,e^-$)$^{97}$Tc$+n$ & Solar $^8$B &  0.3 \\
$^{98}$Mo($\nu,e^-$)$^{97}$Tc$+n$ & SN  &  150 \\
$^{97}$Mo($\nu,e^-$)$^{97}$Tc & Solar $^8$B & 1.2 \\
$^{97}$Mo($\nu,e^-$)$^{97}$Tc & SN  & 18  
\end{tabular}
\end{table}

\begin{table}
\caption{Subgroups in Sco-Cen}
\begin{tabular}{cccc}
Subgroup  & T (My) &  $\langle \Phi \rangle$  (cm$^{-2}$ s$^{-1}$) &
$r_\mathrm{eff}$ (pc) \\
\hline US & 3 & 111 & 154 \\
LCC & 7 & 218  & 110 \\
UCL & 10 & 120 & 148
\end{tabular}
\end{table}

\end{document}